\documentclass[11pt]{article}
\def\D{\Delta}
\def\d{\delta}

\def\l{\lambda}
\def\S{\Sigma}
\def\G{\Gamma}
\def\g{\gamma}
\def\e{\epsilon}
\def\s{\sigma}
\def\o{\omega}

\def\a{\alpha}
\def\b{\beta}

\def\m{\mu}
\def\n{\nu}

\def\s{\sigma}
\def\p{\pi}
\def\f{\phi}
\def\F{\Phi}
\def\th{\theta}
\def\t{\tau}
\def\e{\epsilon}
\def\vf{\varphi}

\def\dc{{\cal D}}
\def\pa{\partial}

\def\det{\textrm{det}}

\def\R{{\bf R}}
\def\bv{{\Big |}}

\newcommand{\be}{\begin{equation}}
\newcommand{\ee}{\end{equation}}
\newcommand{\bea}{\begin{eqnarray}}
\newcommand{\eea}{\end{eqnarray}}
\begin{document}

\begin{center}
\bf{\Large  Piecewise Flat Metrics and Quantum Gravity}
\end{center}

\bigskip
\begin{center}
{\large A. Mikovi\'c} \\
Departamento de Matem\'atica e COPELABS  \\
Universidade Lus\'ofona\\
Av. do Campo Grande, 376, 1749-024 Lisboa, Portugal\\
and\\
Grupo de Fisica Matem\'atica da Universidade de Lisboa\\
Av. Prof. Gama Pinto, 2, 1649-003 Lisboa, Portugal\\
\end{center}

\centerline{E-mail: amikovic@ulusofona.pt}

\bigskip
\bigskip
\centerline{\bf Abstract}
\begin{quotation}
\noindent\small{We introduce a physical piecewise linear metric associated to a Regge triangulation of a smooth 4-manifold. We describe the basic properties of the corresponding geometry in the cases of the Euclidean and the Minkowski signature. In the Minkowski case, we describe the Regge action and how to define the corresponding path integral for the casual triangulations. We also discus the Regge path integral for a  triangulation associated to the Friedman-Lemaitre-Robertson-Walker cosmological model and briefly study the corresponding wavefunctions, namely the Hartle-Hawking and the Vilenkin wavefunction.}\end{quotation}

\newpage
\section{Introduction}

The idea of using manifold triangulations and the corresponding edge lengths to define the path integral for general relativity (GR) was introduced by Tulio Regge and his colaborators in the 1960's, and it was further developed by other researchers, see \cite{RW} for a review and references. However, finding a smooth-manifold limit of the Regge path integral turned out to be an elusive task, and Regge's approach was abandoned by most of the researchers, although in the past 20 years there has been a revival of the quantum Regge calculus in the form of the spin foam models \cite{SF} and in the form of the casual dynamical triangulation models \cite{CDT}. 

The appearence of the superstring theory in the 1980's as a theory of quantum gravity \cite{SS}, introduced the idea that the finitenness of a quantum gravity (QG) theory can be achieved by changing the mathematical structure which describes the spacetime. In the superstring theory case, this was realised by using a loop manifold instead of a manifold. In the 1990's appared the approach of non-commutative geometry \cite{NG}, where the spacetime manifold was replaced by a non-commutative manifold. Each of these approaches has its advantages and disadvantages, and the main problem of the superstring theory is the aparent absence in Nature of the supersymmetry, while in the case of the non-commutative geometry approach, it is a non-geometric nature of the non-commutative manifolds, so that it is difficult to see what is their physical meaning.

Recently a new mathematical structure for the spacetime was proposed, and this is the piecewise linear (PL) manifold structure associated to a smooth manifold triangulation \cite{M,PL}. This is essentially the Regge formulation of GR, but with the assumption that the basic structure of the spacetime is a PL manifold, while a smooth manifold is an approximation valid when the number of space-time cells (4-simplexes) is large and their size is small. This idea is analogous to a situation in hydrodynamics, where a fluid, which is a large collection of molecules, can be approximated as a continuous medium at the scales much larger than the intermolecular distance.

In this paper we are going to introduce a PL metric associated to a triangulation of a smooth manifold. Note that for a manifold triangulation there is a Cayley-Menger metric \cite{CM,H}. However, the CM metric is not dimensionless so that it cannot be a physical metric. The physical PL metric can be introduced as a rescaled CM metric and we will describe the basic properties of the physical PL metric, especially in the case of the Minkowski signature, since in the existing literature, the PL geometry is predominantly discussed in the Euclidean case, see for example \cite{H}. We are also going to define the path integral (PI) in the Minkowski signature case for casual triangulations, since the existing definition \cite{M} applies only to triangulations where all of the edge lengths are spacelike. We will then discus the path integral for triangulations associated to the Friedman-Lemaitre-Robertson-Walker (FLRW) cosmological model, since it is a physical toy model where some key properties of PLQG can be studied.

In section 2 we will describe the Regge calculus, which was originally formulated for the Riemannian manifolds, and we will introduce a physical PL metric. We will also explain what are the problems of the associated path integrals.
In section 3 we will formulate the Regge calculus for a pseudo-Riemannian manifold with a Minkowski signature and define the corresponding physical PL metric, while in section 4 we will define a Regge path integral for a causal triangulation. In section 5 we will define and discus the HH wavefunction for a triangulation corresponding to the FLRW cosmological model, while in section 6 we will do the same for the Vilenkin wavefunction. In section 7 we will present our conclusions.

\section{Regge calculus}

The Regge discretization of GR \cite{R}, amounts to replacing the smooth spacetime manifold $M$ with a simplicial complex $T(M)$ which corresponds to a triangulation of $M$, while the metric on $T(M)$  is determined by the set of the edge lengths
\be\{ L_\e > 0 \,|\, \e \in T(M)\} \,.\label{sel}\ee
Given the edge lengths (\ref{sel}), one would like to define a metric on the PL manifold $T(M)$  such that the PL metric on each 4-simplex $\s$ of $T(M)$ is flat and of the euclidean signature, i.e. $(+,+,+,+)$.

This can be done by using the Cayley-Menger metric \cite{H}
\be G_{\m\n}(\s) = L_{0\m}^2 + L_{0\n}^2 - L_{\m\n}^2 \,,\label{cmm}\ee
where the five vertices of $\s$ are labeled as $0,1,2,3,4$ and $\m,\n = 1,2,3,4$. Although the CM metric is flat in a four-simplex, it is not dimensionless and hence it is not diffeomorphic to $g_{\m\n} = \d_{\m\n}$. This can be remedied by defining a new PL metric
\be g_{\m\n}(\s) = {G_{\m\n}(\s) \over \left(\det G(\s)\right)^{1/4}} \,,\label{eplm}\ee
which is a rescaled CM metric such that $\det\, g_{\m\n} = 1$, provided $\det\, G_{\m\n} > 0$.

Insuring the euclidean signature of a PL metric requires the following restrictions
\bea  \det\, G(\s) &>& 0 \,,\label{er1}\\
 \det\, G(\t) &>& 0 \,,\label{er2}\\
 \det\, G(\D) &>& 0 \,,\label{er3}\eea 
for every 4-simplex $\s$, every tetrahedron $\t$ and every triangle $\D$ of $T(M)$. The last inequality is equivalent to the triangular inequalities for the edge lengths of a triangle. These inequalities permit us to define the volumes of $n$-simplexes via the Cayley-Menger determinants \cite{CM}
\be \det\, G(\s_n ) = 2^n (n!)^2 V^2 (\s_n)\,,\quad n=2,3,4 \,. \ee

Note that for an arbitrary assignment of $L_\e$, the volumes $V_n$ can be positive, zero or imaginary. Just taking the strict triangular inequalities will insure the positivity of the triangle areas, but then some of the higher volumes can be zero or negative. Hence all of the three inequalities must be imposed.

The Einstein-Hilbert (EH) action on $M$ is given by
\be S_{EH} = \int_M \sqrt{\det g}\, R(g) \, d^4 x  \,,\label{eh}\ee
where $R(g)$ is the scalar curvature associated to a metric $g$. On $T(M)$ the EH action becomes the Regge action 
\be S_R (L) = \sum_{\D\in T(M)} A_\D (L)\, \d_\D (L) \,,\label{ra}\ee
where $A_\D = V(\D)$ and the deficit angle $\d_\D$ is given by
\be \d_\D = 2\pi - \sum_{\s \supset \D} \th_\D^{(\s)} \,. \ee
A dihedral angle $\th_\D^{(\s)}$ is defined as the angle between the 4-vector normals associated to the two tetrahedrons that share the triangle $\D$, and it is given by
\be \sin \th_\D^{(\s)} = \frac{4}{3} {A_\D V_\s \over V_\t V_{\t'}} \,. \ee

Given the Regge action (\ref{ra}), the corresponding Euclidean path integral can be written as
\be Z_E = \int_{D} \prod_{\e=1}^{N_1} dL_\e \, \mu(L)\, e^{- S_R (L)/l_P^2} \,,\label{epi}\ee
where $D$ is a subspace of $({\bf R}_+ )^{N_1}$ consistent with the triangular inequalities. One can introduce the path-integral measure $\m (L)$ as 
\be \m (L) = \prod_{\e=1}^{N_1}  \left(L_\e \right)^\a \,, \label{rm}\ee
where $\a$ is a constant, and usually $\a = 1$.

The immediate problem with the choices (\ref{epi}) and (\ref{rm}) is that the finiteness of $Z_E$ is not guaranteed because $S_R(L)$ is not bounded from bellow and the measure (\ref{rm}) does not fall off sufficiently quickly for large $L_\e$ and negative $\a$. A simple way to remedy this is to complexify the Euclidean  path integral via
\be Z_{EC} = \int_{D} \prod_{\e=1}^{N_1}  dL_\e \, \mu(L)\, e^{i S_R (L)/l_P^2} \,.\label{cpi}\ee
However, the problem with (\ref{cpi}) is that it is not clear how to relate it to a path integral for the Minkowski signature metrics, i.e. how to do the Wick rotation.

\section{Minkowski PL metric}

The problems with the euclidean path integrals (\ref{epi}) and (\ref{cpi}) can be avoided by using the Minkowski signature metric from the very beginning. In order to formulate the Regge action in the Minkowski signature case,
we need to discuss certain aspects which are absent in the euclidean case.

The novelty in the Minkowski case is that $L_\e^2$ can be positive or negative, so that $L_\e \in {\bf R}_+$ or $L_\e \in i\, {\bf R}_+$. Consequently we have to indicate in $T(M)$ which edges are space-like (S) and which edges are time-like (T). We will not use the light-like edges ($L_\e^2 = 0$). Although one can triangulate a pseudo-Riemannian manifold such that all the edges are spacelike, it is much simpler and more natural to use the triangulations where we have both the spacelike and the timelike edges.

The CM metric is now given by the same expression as in the euclidean case (\ref{cmm}), while the physical PL metric is given by 
\be g_{\m\n}(\s) = {G_{\m\n}(\s) \over |\det\, G(\s)|^{1/4}} \quad,\ee
where the modulus of the determinant accounts for the fact that now $\det\, G(\s) < 0$.

In order to insure the Minkowski signature of the PL metric we need to impose
\be \det\, G(\s) < 0 \quad, \ee
for any four-simplex $\s$ in $T(M)$. This is analogous to the first restriction in the euclidean case (\ref{er1}). However, there is no need for the analogs of the second (\ref{er2}) and the third restriction (\ref{er3}), since the signatures of $\det\, G(\t)$ and $\det\, G(\D)$ are not fixed in the Minkowski case. Namely, $ \det\, G(\t) > 0$ if $\t$ belongs to a euclidean hyper-plane of $g_{\m\n}(\s)$, while $\det\, G(\t) < 0$ if $\t$ belongs to a Minkowski hyper-plane. Also $\det\, G(\D) > 0$ if $\D$ belongs to a euclidean 
plane while $\det\, G(\D) < 0$ if $\D$ belongs to a Minkowski plane.

The volumes of $n$-simplexes can be defined as
\be (V_n)^2 =  {|\det\, G_n |\over 2^n (n!)^2 } > 0 \,,\quad n = 2,3,4 \,,\ee
so that $V_n > 0$. Note that in the $n=1$ case we should distinguish between the labels $L_\e \in {\bf C}$ and their volumes $|L_\e | > 0$. We will also use an equivalent labeling $L_\e \to |L_\e|$ with an indication $S$ or $T$ for an edge $\e$. 

Given that the edge lengths can take real or imaginary values in a Minkowski space, this implies that the angles between the vectors can be real or complex. Let us consider the angles in a Minkowski plane. Such angles can be defined as
\be \cos\a = {\vec u \cdot \vec v \over ||\vec u||\,||\vec v||}\,,\quad \sin \a = \sqrt{1 -\cos^2 \a}\,, \quad \a\in {\bf C}\,,\label{mia}\ee
where $\vec u = (u_1, u_0)$, $\vec u \cdot \vec v = u_1 v_1 - u_0 v_0 $ and $||\vec u || =\sqrt{ \vec u \cdot \vec u}$. 

Consider two spacelike vectors $\vec u = (1,0)$ and $\vec v = (\cosh a, \sinh a )$, $a\in \bf R$. Since $||\vec u || = ||\vec v || = 1$ then
\be \cos\a = \cosh a \,,\quad \sin\a = i \sinh a \,\Rightarrow \,\a =  i\,a \,.\ee

In the case of a spacelike vector $\vec u = (1,0)$ and a timelike vector $\vec v = (\sinh a, \cosh a )$, we have
$||\vec u || = 1$ and $||\vec v || = i$, so that
\be \cos\a = -i\sinh a \,,\quad \sin\a =  \cosh a \,\Rightarrow \,\a = \frac{\pi}{2} - i\,a \,.\ee

And if we have two timelike vectors $\vec u = (0,1)$ and $\vec v = (\sinh a, \cosh a )$, then
\be \cos\a = \cosh a \,,\quad \sin\a = i \sinh a \,\Rightarrow \,\a =  i\,a \,.\ee

The definition (\ref{mia}) then implies that the sum of the angles between two intersecting lines in a Minkowski plane is $2\pi$.

In order to define the dihedral angles in the Minkowski case we will introduce 
\be (v_n)^2 =  {\det\, G_n \over 2^n (n!)^2 }  \,,\quad n = 2,3,4 \,,\ee
so that $v_n = V_n$  for $\det\,G_n > 0$ or $v_n = i\, V_n$ for $\det\,G_n < 0$. In the $n=1$ case we have $v_\e = L_\e$ for a spacelike edge or $v_\e = iL_\e$ for a timelike edge where $L_\e > 0$. Then the angle between two edges in a triangle is given by
\be \sin \a_\pi^{(\D)} = {2\,v_\D \over v_\e \,v_{\e'}}\,,\label{lda}\ee
where $\pi$ the common point (known as the hinge).

The dihedral angle between two triangles sharing an edge in a tetrahedron is given by
\be  \sin \f_\e^{(\t)} =\frac{3}{2}\, { v_\e \,v_\t \over v_\D \,v_{\D'}}\,,\label{ldb}\ee
while the dihedral angle between two tetrahedrons sharing a triangle in a four-simplex is given by
\be \sin \th_\D^{(\s)} = \frac{4}{3}\,{ v_\D \, v_\s \over v_\t \,v_{\t'}}\quad.\label{ldc}\ee

The formulas (\ref{lda}),(\ref{ldb}) and (\ref{ldc}) are generalizations the corresponding euclidean formulas such that $V_n \to v_n$, and the novelty in the Minkowski case is that $\sin \th$ is not restricted to the interval $[-1,1]$, but $\sin \th \in \bf R$ or $\sin \th \in i{\bf R}$. This also means that the Minkowski dihedral angles can take the complex values.

In the case of a dihedral angle $\th_\D^{(\s)}$ there are two possibilities. If the triangle $\D$ is in a Minkowski (ST) plane, then $\th$ will be an angle in an orthogonal Euclidean (SS) plane, so that
$\sin\th = \sin a$. If $\D$ is in an SS plane, then $\th$ will be in an orthogonal ST plane, so that $\sin\th = \cosh a$ or $\sin\th =i\sinh a$. 

The deficit angle will then take the following values:
\be\d_\D = 2\pi - \sum_{\s\supset\D} \th_\D^{(\s)} \in {\bf R} \,,\ee
when $\D$ is an ST triangle, while
\be\d_\D = 2\pi - \sum_{\s\supset\D} \th_\D^{(\s)} \in  \frac{\pi}{2}{\bf Z} + i{\bf R} \,,\ee
when $\D$ is an SS triangle. Note that for an SS triangle the triangle inequalities are valid, while for an ST triangle they do not apply.

The appearance of the complex values for the deficit angles in the Minkowski signature case raises the question of how to generalize the euclidean Regge action such that the new action is real. A proposal for a lorentzian Regge action was given in \cite{CDT}
\be  \tilde S_R = \sum_{\D \in SS} A_\D \, \frac{1}{i}\,\d_\D + \sum_{\D \in ST} A_\D \,\d_\D  \,.\label{ma} \ee
However, the problem with this definition is that a priori $\tilde S_R \in {\bf R} + i \frac{\pi}{2}{\bf Z}$, so that one has to verify for a given triangulation that $Im\,\tilde S_R = 0$.

In order to avoid this difficulty, we will take
\be \tilde S_R = Re\left(\sum_{\D \in SS} A_\D \, \frac{1}{i}\,\d_\D \right) + \sum_{\D \in ST} A_\D \,\d_\D  \,.\label{mra} \ee
This definition can be justified by the fact that the authors of \cite{CDT} have verified that $Im\,\tilde S_R = 0$ for a special class of triangulations, which are physically relevant, and they are called the casual triangulations.

\section{GR path integral}

In the case of the Minkowski signature PL metrics, the definition of the path integral requires some extra assumptions. Let $M = \S \times [0,n]$, $n\in \bf N$, be a spacetime manifold with two boundaries diffeomorphic to a 3-manifold $\S$. We will use a time-ordered triangulation, which is also known as a causal triangulation \cite{CDT}
\be T(M) = \cup_{k = 0}^{n-1} \, \tilde T_k \left(\S \times [k,k+1]\right) \,, \ee
where $\tilde T_k$ is a triangulation of a slab $\S\times [k,k+1]$ such that 
$$\partial \tilde T_k = T_k (\S) \cup T_{k+1}(\S)$$ 
and $T_k$ are triangulations of $\S$.  We then choose $v_\e = L_\e$ for $\e \in T_k (\S)$ and $v_\e = i L_\e$ for $\e\in \tilde T_k \setminus (T_k \cup T_{k+1})$.

The corresponding path integral can be defined as
\be Z_R = \int_{D} \,\prod_{\e =1}^{N_1} dL_\e \, \mu(L)\, e^{i \tilde{S}_R (L)/l_P^2} \,,\label{rpi}\ee
where $D$ is a region of $({\bf R}_+ )^{N_1}$  consistent with a PL Minkowski geometry\footnote{The easiest way to determine $D$ is to embed $T(M)$ into a 5-dimensional Minkowski space.}. The path-integral measure $\m$ must be chosen such that it makes $Z_R$ convergent. Furthermore, if we want that a quantum effective action $\G (L)$ which can be associated to (\ref{rpi}), becomes $\tilde S_R (L)$ in the classical limit ($L_\e \gg l_P$), then the measure $\m$ has to obey
\be\ln \m (\l\, L_1, \cdots, \l L_N)\approx  O(\l^a)\,,\quad a \ge 2 \,,  \label{asm}\ee
for $\l\to +\infty$, see \cite{M,MV}.  A choice for $\m$ can be made such that it is consistent with (\ref{asm}) and it insures the diffeomorphism invariance of the  smooth-spacetime effective  action \cite{M}. It is given by
\be \m(L) = \exp\left(-V_4 (L)/L_0^4 \right) \,,\label{plm}\ee
where $V_4$ is the volume of $T(M)$ and $L_0$ is a new parameter in the theory. A set of possible values for $L_0$ can be fixed by requiring that the effective cosmological constant coincides with the observed value, see \cite{MV, PL}.

\section{Hartle-Hawking wavefunction}

In canonical quantization of GR  the spacetime $M$ must have the ``slab'' topology $\S\times I$, where the interval $I$ is a subset of $\bf R$, and the spacetime metric is given by
\be ds^2 = - ( N^2 - n^i n_i) dt^2 + 2n_i \,dt dx^i + h_{ij} \,dx^i dx^j \,,\label{cnm}\ee
where $(x^i, t)$ are coordinates on $\S\times I$, $N$ is the lapse and $n^i$ is the shift vector, while $h_{ij}$ is a metric on $\S$. The canonical analysis of the AH action for the metric (\ref{cnm}) gives that the canonical variables are $h_{ij}$ and its canonically conjugate momenta $p_h^{\,\,\,ij}$, which are constrained by the diffeomorphism constraints $D_i(p_h,h)$ and the Hamiltonian constraint $W(p_h ,h)$. The canonical quantization then gives that a wavefunction $\Psi (h)$ has to be invariant under the 3-diffeomorphisms of $\S$ and $\Psi(h)$ has to obey the Wheeler-de Witt (WdW) equation
\be \hat W(\hat p_h , \hat h) \Psi (h) = 0 \,,\label{wdw}\ee
where $\hat W$ is an operator obtained by substituting the variables in the function $W(p_h,h)$ with the operators $\hat{p_h}^{ij}$ and $\hat{h}_{ij}$.

Solving the WdW equation (\ref{wdw}) in general case is notoriously difficult. However, Hartle and Hawking have proposed a way to construct a solution which describes an initial wavefunction of the universe \cite{HH}. It is given by a path integral
\be \Psi_0 (h) = \int \dc g \, e^{-\int_M d^4 x \sqrt{g} \left(R(g)+ \l\right) /l_P^2} \,, \label{hhw}\ee
where the spacetime $M$ has the topology of a cup such that $\partial M = \S$ and the metrics $g$ have the euclidian signature such that $g|_{\pa M} = h$. 

Even the Hartle-Hawking (HH) wavefunction $\Psi_0 (h)$ can be calculated only in some special cases. These are the minisuperspace models where the spacetime metric has a finite number of the degrees of freedom (DOF). For example, the Friedmann-Lemaitre-Robertson-Walker (FLRW) metric is given by
\be ds^2 = -N^2(t)\, dt^2 + a^2 (t)\,( dx^2 + dy^2 + dz^2) \,, \label{flrw}\ee
and there are two DOF, the scaling factor $a$ and the laps $N$.

Consequently
\be  \Psi_0 (a) = \int_J dN \int {\cal D}a \exp \left( -\int_I dt \, L_E (a, \dot a , N)/l_P^2 \right)\,,\label{hhpi} \ee
in the $N(t)=$ constant gauge where $J,I\subseteq \bf R$ and $L_E$ is the euclidean metric extension of the Lagrangian $L(a,\dot a, N)$ corresponding to the metric (\ref{flrw}).

For a general minisuperspace model the path integral (\ref{hhpi}) can be calculated only approximately by using the stationary phase approximation. Also, in order to obtain a solution of the WdW equation the interval $J$ has to be promoted into a contour in the complex plane, see \cite{H1,H2}. Even in the Lorentzian version of (\ref{hhpi}), one has to extend $J$ into a complex plane contour \cite{FLT}.

The PLQG formulation also offers a possibility to calculate the HH wavefunction, since one can mimic the minisuperspace models by using simple triangulations where many of the edge lengths are the same. Also there is an advantage that the spacetime geometry is transparent in the PL case, so that all the domains of integration are uniquely determined and there is no need for complex domains of integration since the convergence is achieved through the PI measure (\ref{plm}). 

For example, let us consider the case when $M = S^4$ (a 4-sphere) such that  $\pa M = \S = S^3$ (a three-sphere). We will then consider a triangulation $T(S^4)= \s_6$ (a 4-dimensional simplicial complex based on six points), which we embed in ${\bf R}^5$,  with $T(S^3)=\s_5$ (a 3-dimensional subcomplex of $\s_6$ based on 5 points), 
such that $L_\e = l > 0$ for $ \e \in \s_5$, $L_\e = s > 0 $ for $\e \in \s_6 \setminus \s_5$.
Consequently
\be S_R (l, s) = {5\sqrt{3}\over 2} \, l^2 \,\d_1 (l,s) + {5\over 2} \, l \sqrt{s^2 - {l^2 \over 4}} \,\d_2 (l,s) \,,\label{emsa}\ee
where
\be  \d_ 1 = 2\pi - 2\a  \,,\quad \d_ 2 = 2\pi - 3\b \,,  \ee
and
\be \sin\a = \frac{\sqrt{s^2 - {3l^2\over 8}}}{\sqrt{s^2 - {l^2\over 3}}}\,,\quad \sin\b = \frac{2\sqrt{2}\sqrt{s^2 - {3l^2 \over 8}}\sqrt{s^2 - {l^2 \over 4}}}{3(s^2 - {l^2 \over 3})}\,.\label{hhda}\ee
 
The HH path integral is then given by
\be  \Psi_0 (l) = \int_{l_1}^{\infty} ds \,\mu (l,s) \,\exp \left( - S_R (l, s)/l_P^2 \right)\,, \label{mshh}\ee
where $l_1 = \sqrt{\frac{3}{8}}\,l$ and
\be \mu (l,s) = \exp\left( - {5\sqrt{2}\,l^3 \over 48\, L_0^4 } \sqrt{s^2 - {3l^2\over 8}}\,  \right) \,.\ee

The integral (\ref{mshh}) is convergent becasue when $s\to +\infty$ we have $\a\to \pi/2$, $\b\to \arcsin(2\sqrt{2}/3)$ so that
\be \m\, e^{-S_R/l_P^2} \approx e^{-\frac{5\sqrt{3}\pi}{4}\, l^2 /l_P^2}\exp\left( - (\l_0 \,l^2 + \d )ls/l_P^2 \right)\,,\ee
where $\l_0 = 5\sqrt{2}\,l_P^2/(48 L_0^4 )$ and $\d = 2\pi - 3\arcsin(2\sqrt{2}/3)$.

The HH integral is also convergent for the trivial measure $\m = 1$, which corresponds to $\l_0 = 0$. However, when a cosmological constant (CC) term is included, the HH integral will be divergent for a sufficiently negative $\l$, since then
\be \m\, e^{-S_R/l_P^2} \approx e^{-\frac{5\sqrt{3}\pi}{4}\, l^2 /l_P^2}\exp\left( - [(\l_0 + \l)l^2 + \d]ls/l_P^2 \right)\,,\ee
and $(\l_0 + \l)l^2 + \d < 0$. This illustrates the earlier observation that a Euclidean GR path integral is not always convergent.

A natural question arises in relation to the PL HH wavefunction (\ref{mshh}), and that is whether it satisfies a WdW equation. Note that the scale factor and the lapse are given by
\be a = l/ l_0 \,,\quad N = s/ t_0 \quad,\ee
where $l_0$ is a unit of length and $t_0$ is a unit of time. One can then ask is there a WdW operator $\hat W$ such that $\hat W \Psi_0 (a) = 0$? More precisely, are there
some constants $\a,\b$ and $\g$ such that
\be\hat W \Psi_0  =  \a \frac{1}{a}{d^2 \Psi \over da^2} + \b\frac{d}{da}\left(\frac{1}{a}{d \Psi \over da}\right) +\g\frac{d^2}{da^2}\left({ \Psi_0 \over a}\right)+ \l\,a^3 \Psi_0 = 0 \,.\label{ewdw}\ee 

The equation (\ref{ewdw}) does not necessarilly holds in PLQG, becase the WdW equation corresponds to a smooth manifold $M$, while we have a PL manifold $T(M)$. However, when $N_1 \to \infty$, i.e. in the smooth-manifold limit, we expect that
\be\hat W_{T(M)} \to \hat W_M \,,\ee
where $\hat W_{T(M)}$ is an operator such that $\hat W_{T(M)} \Psi_0 (a) = 0$.

The bosonic matter can be coupled to gravity by using the euclidean PL metric (\ref{eplm}). In the case of a scalar field $\f$ we have
\be S_m = \sum_{\s\in T(M)} V_\s \,{\cal L}_\s \,,\ee 
where
\be {\cal L}_\s = -\frac{1}{2} g^{\m\n}(\s)\, D_\m \f \,D_\n \f \,-\, U(\f_0) \ee
and $U(\f)$ is the scalar field potential. The metric $g^{\m\n}(\s)$ is given by the inverse matrix of (\ref{eplm}), while
\be D_\m \f = {\f_\m - \f_0 \over L_{0\m}} \,,\ee
where $\f_\m = \f(\xi_\m)$ and $\f_0 = \f(\xi_0)$, while $\xi_0$ and $\xi_\m$ are the verticies of $\s$.

In the case of the triangulation corresponding to the Regge action (\ref{emsa}), we have the following HH wavefuction
\be \Psi(l,f) = \int_{l_1}^{\infty} ds \int_{-\infty}^{\infty} d\vf \,\mu (l,s) \exp \left( -\frac{1}{l_P^2}\left[ S_R (l, s) -  S_m (l,s, \vf , f)\right]\right)\,, \label{mhh}\ee
where $\f (\xi) = f$ for $\xi\in \s_5$ and $\f (\xi) = \vf$ for $\xi\in \s_6 \setminus \s_5$. Note that in the euclidean case the matter action is $-S_m$.

A necessary but not sufficient condition for the convergence of (\ref{mhh}) is that the integral
\be \int_{-\infty}^{\infty} d\vf  \exp \left( \frac{1}{l_P^2}  S_m (l,s, \vf , f)\right)\,, \ee
is convergent. For the polynomial potentials $U(\f)$ this requires that $U(\f) > 0$ for $\f\to\pm\infty$.

The fermionic matter can be coupled by using the PL tetrads $e_\m^a (\s)$, defined by
\be e_\m^a (\s) \, e_\n^b (\s) \,\eta_{ab} = g_{\m\n}(\s) \,, \ee
and the PL spin connection $\o_\m^{ab}(\e^*)$, where $\e^*$ is the dual edge connecting the centers of two adjacent 4-simplexes, see \cite{G}.

\section{Vilenkin wavefunction}

We saw in the previous section that the HH path integral for the PL version of the FLRW minisuperspace model was not always convergent. This is a manifestation of a more general problem that a Euclidean path integral is not always convergent, because, beside the sign of the cosmological constant, the sign of the scalar curvature and the interval of its values affect the convergence of the path integral. One can remedy this situation by using the Minkowski signature metrics, but as we saw in section 4, the definition of a Minkowski path integral requires the spacetime topology of a slab or a cylinder, which is different from the cup topology which was used to define the HH path integral. However, one can simulate the cup topology by collapsing the cylinder to a cone, which is essentially the Vilenkin proposal for the wavefunction of the universe, see \cite{V}.

Note that the path integral (\ref{rpi}) is a function of the initial edge lengths $L_\e = l_\e$ on $T_0(\S)$ and the final edge lengths $L_\e = l'_\e$ on $T_n (\S)$. This function is known as the propagator, which we denote as $G(l,l')$. We denote as $G(h,h')$ the smooth-manifold version of $G(l,l')$, where $h$ is the metric on the final boundary $\S$ and $h'$ is the metric on the initial boundary $\S$. The smooth propagator satisfies a non-homogeneous WdW equation
\be \hat W (\hat p_h , \hat h )\, G(h(x),h'(y)) = \prod_{y\in\S}\d (h(x) - h'(y)) \,.\ee

The Vilenkin proposal for the wavefunction of the universe is to take $h' = 0$, so that $G(h,0)$ satisfies the WdW equation for $h\ne 0$, see \cite{V}. In the PL case, the analog of the Vilenkin wavefunction will be the propagator $G(l,0)$, and it is easy to see that it will be identical to the Lorentzian version of the HH wavefunction for the conical triangulation from the section 5. 

Let us consider 6 points in ${\bf R}^5$ such that
\be v_{jk}^2 = l^2\,,\quad  v_{j5}^2 = 3l^2/8 - t^2 \equiv l_1^2 - t^2
\,, \quad t\in [0,\infty)\,,\label{lint} \ee
where $j,k = 1,2,...,4$ and $j\ne k$. This corresponds to an embedding of $\s_6$ into $\R^5$ with coordinates $(x,y,z,w,t)$ such that $\s_5$ is embedded into the spatial hyperplane $t=0$ with the point $(0,0,0,0,0)$ corresponding to the center of the 3-sphere which contains the spatial points $(x_j,y_j,z_j,w_j,0)$ such that $d_{ij} = l$.

Note that 
\be v_{j5} = \left\{ \begin{array}{lr} s & t\le l_1 \\  is & t > l_1, \end{array} \right. \ee
where
\be s = \sqrt{|l_1^2 - t^2|} \ee
and $l_1 = \sqrt{3l/8}$ is the radius of the 3-sphere.

We will then define the lorentzian HH integral as
\be \Psi_{0}(l) = \int_0^{l_1} ds \,\m(s,l)\, e^{i (\tilde S_R (s,\,l)+\l V_4 ) /l_P^2} + \int_0^{\infty} ds \,\m(s,l)\, e^{i (\tilde S_R (s,\,l)+ \l V_4 )/l_P^2} \,,\label{lhh}\ee
where
\be S_R (l, s) = {5\sqrt{3}\over 2} \, l^2 \,\d_1 (l,s) + {5\over 2} \, l \sqrt{\bv s^2 - {l^2 \over 4}\bv} \,\d_2 (l,s)  \ee
and $\tilde S_R$ is given by (\ref{mra}), while
\be  \d_ 1 = 2\p -2 \a  \,,\quad \d_ 2 = 2\pi - 3\b \,,  \ee
and
\be V_4 (l,s) = {5\sqrt{2}\over 48}\, l^3 \,\sqrt{\bv s^2 - \frac{3}{8}\,l^2 \bv}\,. \ee

The expressions for $\a$ and $\b$ will be given by the definition (\ref{ldc}) and by the embedding (\ref{lint}). In order to see what is $\tilde S_R$, it is helpful to write the dihedral angles in terms of the $t$ variable
\be \sin\a = \left\{ \begin{array}{lr} {it\over \sqrt{l_3^2 - t^2}}=i\sinh a \,, & t < l_3 \\ {t\over \sqrt{t^2 - l_3^2}}= \cosh a\,, & t > l_3  \end{array} \right. \ee
and
\be \sin\b = \left\{ \begin{array}{lr} \frac{2\sqrt 2}{ 3}\,{it\sqrt{l_2^2 - t^2}\over l_3^2 - t^2}=i\sinh b \,, & t < l_3 \\ \frac{2\sqrt 2}{ 3}\,{t\sqrt{ t^2 - l_2^2}\over t^2 - l_3^2}=\sin b \,, & t > l_3  \end{array} \right.\,, \ee
where $l_2 = l/\sqrt{8}$ and $l_3 = l/\sqrt{24}$.

Note that these expressions can be obtained from the Euclidean expressions (\ref{hhda}) by performing an analytic continuation $t \to it$ (a Wick rotation).

The convergence of the lorentzian HH integrals (\ref{lhh}) reduces to the convergence of the second integral. This integral is convergent due to the large-$s$ asymptotics
\be \m\, e^{i S_R} \approx C\, e^{-\l_0 l^3 s/l_P^2}\,e^{ i(\l l^2 + \d) l s/l_P^2} \,,\ee
and due to the asymptotics for $t\to l_3^{\pm 0}$
\be a \approx -\frac{1}{2} \ln |t - l_3| \,,\quad b \approx - \ln |t - l_3| \,.\ee

\section{Conclusions}

Note that the Vilenkin PL wavefunction (\ref{lhh}) is defined for any value of the cosmological constant $\l$, while the Hartle-Hawking PL wavefunction (\ref{hhpi}) is defined only for 
\be\l > -\l_0 - \frac{\d}{l^2} \,.\ee
This reflects the fact that the Lorentzian path integral (\ref{rpi}) has better convergence properties than the Euclidean path integral (\ref{epi}). Also, the Wilenkin PL wavefunction is the same as the lorentzian version of the HH PL wavefunction, a fact which indicates that this may be true in the smooth limit approximation. 

Note that the propagator $G(h,h')$ is not the same as the Schrodinger equation propagator $K(\tilde h,T; \tilde h',T')$, where $K$ satisfies the deparametrized form of the WdW equation
\be \left[i\hbar {\pa\over\pa T} - \hat H ( p_{\tilde h}, \tilde h, T)\right] K(\tilde h,T; \tilde h',T')= 0 \,.\label{sqg}\ee

Namely, one can perform a deparametrization of the Hamiltonian constraint $W(p_h,h)$, which amounts to performing a canonical transformation $(p_h,h) \to ( p_{\tilde h}, P_T ; \tilde h , T)$ such that
\be W(p_h,h) = 0 \Leftrightarrow P_T + H(p_{\tilde h}, \tilde h , T) =0 \,.\ee
Consequently, the canonical quantization with respect to the new canonical variables will give the Schrodinger equation (\ref{sqg}) in the gauge $\pa_i T = 0$, i.e. a gauge choice where $T$ does not depend on the spatial coordinates $x^i$ \footnote{Existence of such a gauge choice is called the problem of time in canonical GR, see \cite{pt}.}.

For example, in the case of the relativistic particle, when the WdW equation becomes the KG equation, we have for the propagator
\be G(\vec x,t) = \int_{\R^4} d\o \, d^3 \vec k \, {e^{-i\o t + i\vec{k}\cdot \vec x} \over \o^2 - {\vec k}^2 - \o_0^2} \,,\ee
where $\o = p_0/\hbar$, $\vec k = \vec p /\hbar$ and $\o_0 = m/\hbar$, while for the Schrodinger propagator one obtains
\be K(\vec x,t) = \int_{\R^3} d^3 \vec k \,e^{-i t\sqrt{{\vec k}^2 + \o_0^2} + i\vec k \cdot \vec x} \,. \ee

In the PL context, the Schrodinger propagator becomes $K(\tilde l_\e, T_k ; \tilde l'_\e, T_{k'})$, where $T_k = k l_0$ is an PL analogue of the time variable $T$ and $k,k' \in \bf N$. The choice of the time variable can be implemented through the following restrictions
\be V(\S_k) = f(k)\, l_0^3 \,,\quad k=0,1,2,... ,n\,, \label{tvg}\ee
where $f(k)$ is a given function. The restrictions (\ref{tvg})  have to be imposed in the path integral (\ref{rpi}) in order to obtain the Schrodinger propagator $K$. 

Note that any solution of a deparametrized WdW equation can be written as
\be \Phi (\tilde h, T) = \int {\cal D}\tilde h' \, K(\tilde h, T; \tilde h', 0)\,\Phi_0 (\tilde h') \,, \ee
where $\Phi_0 (\tilde h')$ is the initial wavefunction of the universe (WFU). The functional  $\F_0 (\tilde h)$ is arbitrary, in contrast to the HH or the Vilenkin vawefunctions, which should satisfy the WdW equation $\hat W \Psi (h)=0$. It looks like that there is a greater freedom in choosing a WFU in the Schrodinger framework than in the WdW framework, so that this question deserves a further study.

The study of the HH path integral with matter (\ref{mhh}) and its lorentzian version are obvious further steps in the PL approach to the problem of the WFU.

Another interesting problem for a further study would be a determination of the smooth limits of the PL Hartle-Hawking and the Vilenkin wavefunctions which can be done by using a conical spacetime triangulation where the spatial simplicial complex $\s_5$ is replaced by $\s_n$ and then let $n\to \infty$.

\end{document}